\documentclass{pasj00}
\usepackage{comment}
\usepackage{ifthen}

\newcommand{\forloop}[5][1]%
{%
\setcounter{#2}{#3}%
\ifthenelse{#4}%
	{%
	#5%
	\addtocounter{#2}{#1}%
	\forloop[#1]{#2}{\value{#2}}{#4}{#5}%
	}%
	{%
	}%
}%

\newcommand{\ctbd}[1]{}

\newcommand{\lc}{light curve}
\newcommand{\lcs}{light curves}
\newcommand{\Lc}{Light curve}

\newcommand{\band}[1]{\ensuremath{#1}~band}

\newcommand{\kms}{\ensuremath{\rm km\,s^{-1}}}
\newcommand{\ms}{\ensuremath{\rm m\,s^{-1}}}

\newcommand{\gcmc}{\ensuremath{\rm g\,cm^{-3}}}
\newcommand{\ergscmsq}{\ensuremath{\rm erg\,s^{-1}\,cm^{-2}}}

\newcommand{\vsini}{\ensuremath{v \sin{i}}}
\newcommand{\feh}{\ensuremath{\rm [Fe/H]}}

\newcommand{\vmac}{\ensuremath{v_{\rm mac}}}
\newcommand{\vmic}{\ensuremath{v_{\rm mic}}}

\newcommand{\rsun}{\ensuremath{R_{\solar}}}
\newcommand{\msun}{\ensuremath{M_{\solar}}}
\newcommand{\lsun}{\ensuremath{L_{\solar}}}

\newcommand{\rstar}{\ensuremath{R_\star}}
\newcommand{\mstar}{\ensuremath{M_\star}}
\newcommand{\lstar}{\ensuremath{L_\star}}

\newcommand{\teffstar}{\ensuremath{T_{\rm eff\star}}}

\newcommand{\loggstar}{\ensuremath{\log{g_{\star}}}}

\newcommand{\rpl}{\ensuremath{R_{p}}}
\newcommand{\mpl}{\ensuremath{M_{p}}}

\newcommand{\rhopl}{\ensuremath{\rho_{p}}}

\newcommand{\arstar}{\ensuremath{a/\rstar}}
\newcommand{\zrstar}{\ensuremath{\zeta/\rstar}}

\newcommand{\rjup}{\ensuremath{R_{\rm J}}}
\newcommand{\mjup}{\ensuremath{M_{\rm J}}}

\newcommand{\refsec}[1]{\mbox{\S\ \ref{sec:#1}}}

\newcommand{\reffigl}[1]{Figure~\ref{fig:#1}}
\newcommand{\refsecl}[1]{\mbox{Section \ref{sec:#1}}}

\newcommand{\reftabl}[1]{Table~\ref{tab:#1}}

\newcommand{\flwof}{\mbox{FLWO 1.2\,m}}

\newcommand{\hatcurCCra}{\ensuremath{02^{\mathrm h}21^{\mathrm m}31.97{\mathrm s}}}                                  
\newcommand{\hatcurCCdec}{\ensuremath{+32{\arcdeg}14{\arcmin}46.1{\arcsec}}}                                 
\newcommand{\hatcurCCmag}{12.56}                                      
\newcommand{\hatcurCCtwomass}{2MASS~02213197+3214461}                  
\newcommand{\hatcurCCgsc}{GSC~2314-00559}                              
\newcommand{\hatcurCCtassmv}{\ensuremath{12.557\pm0.094}}                                   
\newcommand{\hatcurCCtwomassJmag}{\ensuremath{10.989\pm0.022}}         
\newcommand{\hatcurCCtwomassHmag}{\ensuremath{10.593\pm0.021}}         
\newcommand{\hatcurCCtwomassKmag}{\ensuremath{10.501\pm0.017}}         
\newcommand{\hatcurLCrprstar}{\ensuremath{0.0918\pm0.0016}}            
\newcommand{\hatcurLCbsq}{\ensuremath{0.139_{-0.072}^{+0.083}}}        
\newcommand{\hatcurLCimp}{\ensuremath{0.372_{-0.132}^{+0.094}}}        
\newcommand{\hatcurLCzeta}{\ensuremath{17.46\pm0.12}}                  
\newcommand{\hatcurLCdur}{\ensuremath{0.1267\pm0.0015}}                
\newcommand{\hatcurLCingdur}{\ensuremath{0.0122\pm0.0013}}             
\newcommand{\hatcurLCP}{\ensuremath{4.640382\pm0.000032}}              
\newcommand{\hatcurLCPprec}{\ensuremath{4.640382}}                    
\newcommand{\hatcurLCPshort}{\ensuremath{4.6404}}                      
\newcommand{\hatcurLCT}{\ensuremath{2455863.11957\pm0.00035}}          
\newcommand{\hatcurSMEiteff}{\ensuremath{5330\pm100}}                  
\newcommand{\hatcurSMEizfeh}{\ensuremath{+0.06\pm0.1}}                  
\newcommand{\hatcurSMEizfehshort}{\ensuremath{+0.06}}                   
\newcommand{\hatcurSMEilogg}{\ensuremath{4.42\pm0.1}}                  
\newcommand{\hatcurSMEivsin}{\ensuremath{0.4\pm0.5}}                   
\newcommand{\hatcurSMEivmac}{\ensuremath{3.33}}                        
\newcommand{\hatcurSMEivmic}{\ensuremath{0.85}}                        
\newcommand{\hatcurSMEiiteff}{\ensuremath{5330\pm100}}                 
\newcommand{\hatcurSMEiizfeh}{\ensuremath{+0.06\pm0.10}}                
\newcommand{\hatcurSMEiizfehshort}{\ensuremath{+0.06}}                  
\newcommand{\hatcurSMEiilogg}{\ensuremath{4.42\pm0.1}}                 
\newcommand{\hatcurSMEiivsin}{\ensuremath{0.4\pm0.5}}                  
\newcommand{\hatcurSMEiivmac}{\ensuremath{3.33}}                       
\newcommand{\hatcurSMEiivmic}{\ensuremath{0.85}}                       
\newcommand{\hatcurTRESgamma}{\ensuremath{-19.7\pm0.1}}                
\newcommand{\hatcurLBii}{\ensuremath{0.3464}}                          
\newcommand{\hatcurLBiii}{\ensuremath{0.2857}}                         
\newcommand{\hatcurISOmshort}{\ensuremath{0.89}}                       
\newcommand{\hatcurISOmlong}{\ensuremath{0.886\pm0.044}}               
\newcommand{\hatcurISOrshort}{\ensuremath{0.92}}                       
\newcommand{\hatcurISOrlong}{\ensuremath{0.923_{-0.067}^{+0.096}}}     
\newcommand{\hatcurISOlogg}{\ensuremath{4.45\pm0.08}}                  
\newcommand{\hatcurISOlum}{\ensuremath{0.62_{-0.10}^{+0.16}}}          
\newcommand{\hatcurISOmv}{\ensuremath{5.44\pm0.24}}                    
\newcommand{\hatcurISOage}{\ensuremath{10.1\pm4.8}}                    
\newcommand{\hatcurISOMK}{\ensuremath{3.56\pm0.20}}                    
\newcommand{\hatcurRVK}{\ensuremath{35.4\pm2.4}}                       
\newcommand{\hatcurRVk}{\ensuremath{0.003\pm0.028}}                    
\newcommand{\hatcurRVh}{\ensuremath{-0.018\pm0.083}}                   
\newcommand{\hatcurRVfitrms}{\ensuremath{5.7}}                         
\newcommand{\hatcurRVeccen}{\ensuremath{0.067\pm0.047}}                
\newcommand{\hatcurRVomega}{\ensuremath{240\pm104}}                    
\newcommand{\hatcurPPi}{\ensuremath{88.3\pm0.7}}                       
\newcommand{\hatcurPPlogg}{\ensuremath{2.99\pm0.08}}                   
\newcommand{\hatcurPPar}{\ensuremath{12.17\pm1.04}}                    
\newcommand{\hatcurPParel}{\ensuremath{0.0523\pm0.0009}}               
\newcommand{\hatcurPPrho}{\ensuremath{0.59\pm0.16}}                    
\newcommand{\hatcurPPmshort}{\ensuremath{0.27}}                        
\newcommand{\hatcurPPmlong}{\ensuremath{0.267\pm0.020}}                
\newcommand{\hatcurPPrshort}{\ensuremath{0.82}}                        
\newcommand{\hatcurPPrlong}{\ensuremath{0.825_{-0.063}^{+0.092}}}      
\newcommand{\hatcurPPmrcorr}{\ensuremath{0.18}}                        
\newcommand{\hatcurPPteff}{\ensuremath{1082\pm55}}                     
\newcommand{\hatcurPPtheta}{\ensuremath{0.038\pm0.004}}                
\newcommand{\hatcurPPfluxavg}{\ensuremath{3.09_{-0.50}^{+0.80}}}       
\newcommand{\hatcurPPfluxavgdim}{\ensuremath{8}}                       
\newcommand{\hatcurXsecondary}{\ensuremath{2455865.449\pm0.083}}       
\newcommand{\hatcurXsecdur}{\ensuremath{0.1229\pm0.0176}}              
\newcommand{\hatcurXsecingdur}{\ensuremath{0.0117\pm0.0029}}           
\newcommand{\hatcurXdist}{\ensuremath{249_{-19}^{+26}}}                

\newcommand{\hatcur}{HAT-P-38}
\newcommand{\hatcurb}{HAT-P-38b}

\newcommand{\hatcurCCtassvi}{\ensuremath{0.87\pm0.13}}                  
\newcommand{\hatcurSMEversion}{i}                                       
\newcommand{\hatcurSMEteff}{\ifthenelse{\equal{\hatcurSMEversion}{i}}{\hatcurSMEiteff}{\hatcurSMEiiteff}}
\newcommand{\hatcurSMEzfeh}{\ifthenelse{\equal{\hatcurSMEversion}{i}}{\hatcurSMEizfeh}{\hatcurSMEiizfeh}}
\newcommand{\hatcurSMEzfehshort}{\ifthenelse{\equal{\hatcurSMEversion}{i}}{\hatcurSMEizfehshort}{\hatcurSMEiizfehshort}}
\newcommand{\hatcurSMElogg}{\ifthenelse{\equal{\hatcurSMEversion}{i}}{\hatcurSMEilogg}{\hatcurSMEiilogg}}
\newcommand{\hatcurSMEvsin}{\ifthenelse{\equal{\hatcurSMEversion}{i}}{\hatcurSMEivsin}{\hatcurSMEiivsin}}
\newcommand{\hatcurSMEvmac}{\ifthenelse{\equal{\hatcurSMEversion}{i}}{\hatcurSMEivmac}{\hatcurSMEiivmac}}
\newcommand{\hatcurSMEvmic}{\ifthenelse{\equal{\hatcurSMEversion}{i}}{\hatcurSMEivmic}{\hatcurSMEiivmic}}

\newcommand{\hatcurisoshort}{YY}
\newcommand{\hatcurisofull}{Yonsei-Yale (YY)}
\newcommand{\hatcurisocite}{yi:2001}
\newcommand{\hatcurlumind}{\arstar}
\newcommand{\hatcurjhkfilset}{ESO}

\newboolean{rvtablelong}
\setboolean{rvtablelong}{true}

\begin{document}
\SetRunningHead{Sato et al.}{\hatcurb{}}

\title{\hatcurb{}: A Saturn-Mass Planet Transiting a Late G Star}

%

%
 \author{%
   Bun'ei \textsc{Sato}\altaffilmark{1}
   Joel~D.\ \textsc{Hartman}\altaffilmark{2,3}
   G\'asp\'ar~\'A.\ \textsc{Bakos}\altaffilmark{2,3}
   Bence~\textsc{B\'eky}\altaffilmark{3}
   Guillermo~\textsc{Torres}\altaffilmark{3}
   David~W.\ \textsc{Latham}\altaffilmark{3}
   G\'eza~\textsc{Kov\'acs}\altaffilmark{4}
   Zolt\'an~\textsc{Csubry}\altaffilmark{2,3}
   Kaloyan~\textsc{Penev}\altaffilmark{2,3}
   Robert~W.\ \textsc{Noyes}\altaffilmark{3}
   Lars~A.\ \textsc{Buchhave}\altaffilmark{5}
   Samuel~N.\ \textsc{Quinn}\altaffilmark{3,6}
   Mark~\textsc{Everett}\altaffilmark{3}
   Gilbert~A.\ \textsc{Esquerdo}\altaffilmark{3}
   Debra~A.\ \textsc{Fischer}\altaffilmark{7}
   Andrew~W.\ \textsc{Howard}\altaffilmark{8}
   John~A.\ \textsc{Johnson}\altaffilmark{9}
   Geoff~W.\ \textsc{Marcy}\altaffilmark{8}
   Dimitar~D.\ \textsc{Sasselov}\altaffilmark{3}
   Tam\'as~\textsc{Szklen\'ar}\altaffilmark{3} 
   J\'ozsef \textsc{L\'az\'ar}\altaffilmark{10}
   Istv\'an \textsc{Papp}\altaffilmark{10}
   P\'al \textsc{S\'ari}\altaffilmark{10}}

 \altaffiltext{1}{Department of Earth and Planetary Sciences,
Tokyo Institute of Technology, Tokyo, Japan}
 \email{satobn@geo.titech.ac.jp}
 \altaffiltext{2}{Department of Astrophysical Sciences, Princeton University, Princeton, NJ, USA}
 \altaffiltext{3}{Harvard-Smithsonian Center for Astrophysics, Cambridge, MA, USA}
 \altaffiltext{4}{Konkoly Observatory, Budapest, Hungary}
 \altaffiltext{5}{Niels Bohr Institute, University of Copenhagen, DK-2100, Denmark, and Centre for Star and Planet Formation, Natural History Museum of Denmark, DK-1350 Copenhagen}
 \altaffiltext{6}{Department of Physics and Astronomy, Georgia State University, Atlanta, GA, USA}
 \altaffiltext{7}{Astronomy Department, Yale University, New Haven, CT}
 \altaffiltext{8}{Department of Astronomy, University of California, Berkeley, CA}
 \altaffiltext{9}{California Institute of Technology, Department of Astrophysics, MC~249-17, Pasadena, CA}
 \altaffiltext{10}{Hungarian Astronomical Association}


\maketitle

\begin{abstract}
We report the discovery of \hatcurb{}, a Saturn-mass exoplanet
transiting the V=\hatcurCCmag\ dwarf star \hatcurCCgsc\ on a
$P=\hatcurLCPshort$\,d circular orbit.  The host star is a
\hatcurISOmshort\,\msun{} late G-dwarf, with solar metallicity, and a
radius of \hatcurISOrshort\,\rsun.  The planetary companion has a mass
of \hatcurPPmshort\,\mjup{}, and radius of \hatcurPPrshort\,\rjup.
\hatcurb{} is one of the closest planets in mass and radius to Saturn
ever discovered.
\end{abstract}

\section{Introduction}
\label{sec:introduction}

Transiting extrasolar planets (TEPs) provide unique opportunities
to study the bulk physical properties (mass, radius, mean density)
of planetary mass objects. These data can be compared with
theoretical predictions based on planet interior models to give us 
insight into the formation and evolution of planets
(e.g. \cite{baraffe:2008,fortney:2007,burrows:2007,seager:2007}).
To date, more than 150 TEPs identified by dedicated photometric surveys
have been confirmed by dynamical mass
determinations\footnote{e.g. http://exoplanets.org},
and the planets have been found to show great
diversity in their properties. For example, at the high-mass end
Jupiter-mass and super-Jupiter-mass planets ($M\gtrsim 0.4~\mjup$),
which are most likely H/He-dominated gas giants, exhibit a wide range
of radii and inferred core masses (e.g. \cite{cabrera:2010,anderson:2011}).
At the low-mass end
($M\lesssim 0.1~\mjup$), super-Earths and Neptunes may have diversity
in their bulk compositions (e.g. \cite{valencia:2010,charbonneau:2009,
lissauer:2011}).

In between, the sample size of known Saturn-mass TEPs ($0.1 < M < 0.4
M_J$)\footnote{There is no strict definition of Saturn-mass planets.
The mass range is somewhat arbitrarily set limit for these lighter TEPs.}
is still quite small probably because these planets are
intrinsically rare at short orbital periods (\cite{hartman:2009}).
However, the 11 previously known Saturn-mass TEPs also appear to show
diverse properties, like the more massive planets, with a wide range
of mean densities, from 0.3 to 2.3 times that of Saturn
($\rhopl=0.19~\gcmc$ for WASP-39b and $\rhopl=1.6~\gcmc$ for CoRoT-8b,
respectively; \cite{faedi:2011,borde:2010}).  A correlation
between metallicity and inferred core mass
(e.g. \cite{hartman:2009,miller:2011}), and also between metallicity,
irradiation (planetary equilibrium temperature), and planet radii has
been suggested for the Saturn-mass TEPs
(e.g. \cite{enoch:2011,beky:2011}), but it should be confirmed by
collecting a larger number of samples.

The Hungarian-made Automated Telescope Network (HATNet;
\cite{bakos:2004}) survey for transiting exoplanets (TEPs) around
bright stars ($9\lesssim r \lesssim 14.5$) operates six wide-field
instruments: four at the Fred Lawrence Whipple Observatory (FLWO) in
Arizona, and two on the roof of the hangar servicing the Smithsonian
Astrophysical Observatory's Submillimeter Array, in Hawaii.  Since
2006, HATNet has announced and published 37 TEPs
(e.g.~\cite{hartman:2011}).  In this work we report our thirty-eighth
discovery, around the relatively bright star \hatcurCCgsc{}.
\hatcurb{} is one of the closest planets in mass and radius to Saturn
ever discovered.

In \refsecl{obs} we summarize the detection of the photometric transit
signal and the subsequent spectroscopic and photometric observations
of \hatcur{} to confirm the planet.  In \refsecl{analysis} we analyze
the data to rule out false positive scenarios, and to determine the
stellar and planetary parameters. Our findings are discussed in
\refsecl{discussion}.

\section{Observations}
\label{sec:obs}

The observational procedure employed by HATNet to discover TEPs has
been described in detail in several previous discovery papers
(e.g. \cite{bakos:2010}, \cite{latham:2009}). In the following
subsections we highlight specific details that are pertinent to the
discovery of \hatcurb{}.

\subsection{Photometric detection}
\label{sec:detection}

The transits of \hatcurb{} were detected with the HAT-5 telescope in
Arizona and the HAT-8 telescope in Hawaii.  A total of 2381 exposures
of 5 minutes and 6126 exposures of 3 minutes were obtained in Sloan
$r$ band for a $\sim
10\degree \times 10\degree$ field containing \hatcurCCgsc{} between
August and November 2010. We reduced these observations to light
curves for the 36,000 stars in the field with $r \lesssim 14.5$ whose
positions were derived from
the Two Micron All Sky Suvey catalog (2MASS; \cite{skrutskie:2006})
using a custom image subtraction (\cite{alard:2000}) based procedure
with discrete kernels (\cite{bramich:2008}), as described in
\citet{pal:2009}. After decorrelating against external parameters and
detrending with the Trend Filtering Algorithm (TFA;
\cite{kovacs:2005}), we achieved a per-image photometric precision of
3.5\,mmag for the brightest stars in the field.

\begin{figure}[!ht]
  \begin{center}
    \FigureFile(80mm,56mm){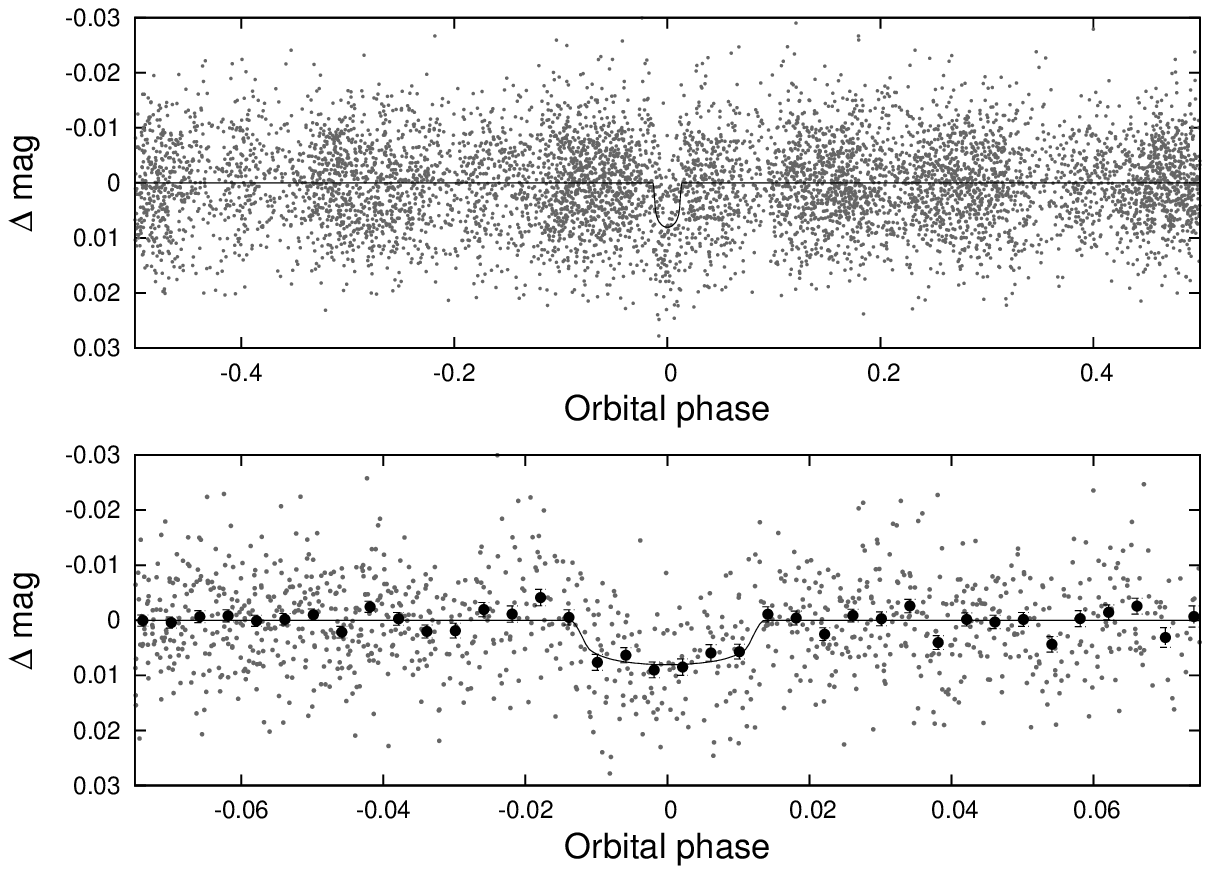}
  \end{center}
  \caption{
    Unbinned \lc{} of \hatcur{} including all $\sim 8500$ instrumental
    \band{r} measurements obtained with the HAT-5 and HAT-8 telescopes
    of HATNet (see the text for details), and folded with the period
    $P = \hatcurLCPprec$\,days resulting from the global fit described
    in \refsecl{analysis}.  The solid line shows a simplified transit
    model fit to the light curve (\refsecl{globmod}). The bottom panel
    shows a zoomed-in view of the transit, the filled circles show the
    light curve binned in phase with a bin-size of 0.004.
}\label{fig:hatnet}
\end{figure}

The \lcs{} were searched for transits using
the Box Least-Squares (BLS; \cite{kovacs:2002}) method.  We detected
a significant signal in the \lc{} of \hatcurCCgsc{} (also known as
\hatcurCCtwomass{}; $\alpha = \hatcurCCra$, $\delta = \hatcurCCdec$;
J2000; V=\hatcurCCtassmv\, \cite{droege:2006}), with 
a period of $P=\hatcurLCPshort$\,days
(see \reffigl{hatnet}).  

\subsection{Reconnaissance Spectroscopy}
\label{sec:recspec}

High-resolution, low-S/N reconnaissance spectra were obtained for
\hatcur{} using the Tillinghast Reflector Echelle Spectrograph (TRES;
\cite{furesz:2008}) on the 1.5\,m Tillinghast Reflector at FLWO. Three
TRES spectra of \hatcur{} were obtained, reduced and analyzed to
measure the stellar effective temperature, surface gravity, projected
rotation velocity, and radial velocity (RV)
via cross-correlation against a library of
synthetic template spectra. For details see \citet{quinn:2012} and
\citet{buchhave:2010}. The resulting measurements are given in
\reftabl{reconspec}.

These observations revealed no detectable RV variation at the
$\sim 0.1$\,\kms\ precision of the observations. Additionally the
spectra are consistent with a single, slowly-rotating, dwarf star.

\begin{longtable}{lrrrr}
  \caption{Reconnaissance Spectroscopy of \hatcur{} with TRES.}\label{tab:reconspec}
  \hline              

  \noalign{\vskip 2pt}
  \multicolumn{1}{c}{HJD} & 
  \multicolumn{1}{c}{T$_{\rm eff}$} & 
  \multicolumn{1}{c}{$\log g$} &
  \multicolumn{1}{c}{$\vsini$} & 
  \multicolumn{1}{c}{RV} \\
  \multicolumn{1}{c}{\hbox{(2,455,000$+$)}} &
  \multicolumn{1}{c}{(K)} &
  \multicolumn{1}{c}{(cgs)} &
  \multicolumn{1}{c}{(\kms)} &
  \multicolumn{1}{c}{(\kms)} \\
  [1.0ex]
  \endfirsthead
  \hline
\endhead
  \hline
\endfoot
  \hline
\multicolumn{5}{p{7cm}}{\footnotesize Uncertainties on T$_{\rm eff}$, $\log g$, $\vsini$, and the RV are $\sim 100$\,K, $0.1$\,dex, $0.5$\,\kms, and $\sim 0.1$\,\kms, respectively. The stellar parameters were derived with the assumption of $\feh=0.0$.} \\
\endlastfoot
  \hline
    \noalign{\vskip 2pt}
546.66915 & 5510 & 4.57 & 2.4 & $-$19.60 \\
576.59133 & 5460 & 4.49 & 2.5 & $-$19.83 \\
578.64318 & 5420 & 4.39 & 2.4 & $-$19.72 \\
    [1.0ex] 
\end{longtable}

\subsection{High resolution, high S/N spectroscopy}
\label{sec:hispec}

\begin{figure}[!ht]
  \begin{center}
    \FigureFile(80mm,118mm){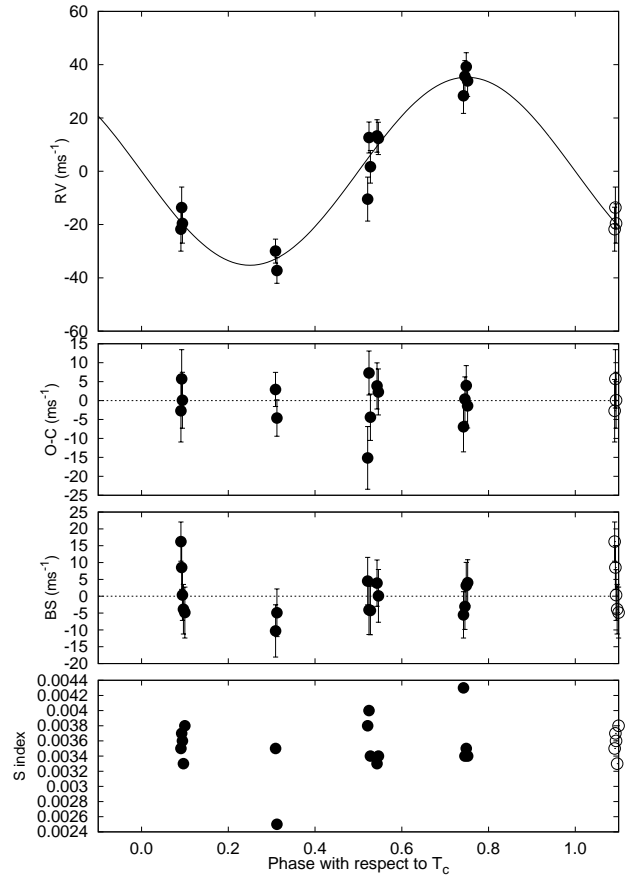}
  \end{center}
  \caption{
	{\em Top panel:} Subaru/HDS RV measurements for
        \hbox{\hatcur{}} shown as a function of orbital phase, along
        with our best-fit model (see \reftabl{planetparam}).  Zero
        phase corresponds to the time of mid-transit.  The
        center-of-mass velocity has been subtracted.
	{\em Second panel:} Velocity $O\!-\!C$ residuals from the best
        fit.
	{\em Third panel:} Bisector spans (BS) for the HDS spectra,
        with the mean value subtracted.  The measurement from the
        template spectrum is included (see \refsecl{hispec}).
	{\em Bottom panel:} Relative chromospheric activity index $S$
        measured from the HDS spectra.
	Note the different vertical scales of the panels. Observations
        shown twice are represented with open circles.
}\label{fig:rvbis}
\end{figure}

We proceeded with the follow-up of this candidate by obtaining
high-resolution, high-S/N spectra to detect and characterize RV
variations induced by the putative planet.  For this we used the
High-Dispersion Spectrograph (HDS; \cite{noguchi:2002}) with the
I$_{2}$ absorption cell (\cite{kambe:2002}) on the Subaru telescope.
We obtained a total of 16 spectra, including two I$_{2}$-free
observations which we used to construct a template for the RV
measurements. The observations were made on four nights between 2011
August 05 and 2011 August 08 using the KV370 filter, a $0\farcs6
\times 2\farcs0$ slit, giving a wavelength resolution
($\lambda/\Delta\lambda$) of 60000, and the StdI2b setup, covering a
wavelength region of 3500--6200~\AA. We used exposure times of 10 to
20 minutes, yielding a S/N per resolution element of 45 to 80 at
$5500$\,\AA, depending on seeing. The spectra were extracted and
reduced to relative RVs in the solar system barycentric frame
following the procedure of \citet{sato:2002} and \citet{sato:2005},
which is based
on the method by \citet{butler:1996}. In this technique a star+I$_2$
spectrum is modeled as a product of a high resolution I$_2$ and a
stellar template spectrum convolved with a modeled instrumental
profile (IP) of the spectrograph. To obtain the stellar template,
\citet{sato:2002} extracted a high resolution stellar spectrum from
several star+I$_2$ spectra. However, we have since found that when
applying this technique for obtaining a template for the HDS spectra
systematic errors sometimes appear in the resulting RVs. These errors
disappear when when we use a stellar template obtained by deconvolving
a pure stellar spectrum with the spectrograph IP estimated from a
B-star+I$_2$ spectrum. We applied this latter technique to generate
the template for our HDS observations of \hatcur{}. The resulting RV
measurements and their uncertainties are listed in \reftabl{rvs}. The
period-folded data, along with our best fit described below in
\refsecl{analysis}, are displayed in \reffigl{rvbis}.

In addition to the HDS/Subaru observations, we also obtained a single
high-resolution, high-S/N spectrum using the HIRES instrument
(\cite{vogt:1994}) on the Keck~I telescope. This observation was made
without the I$_{2}$ absorption cell on the night of 2011 February 19,
and was used to determine the stellar parameters as described in
\refsecl{stelparam}.

A bisector analysis based on the HDS spectra was done following a
procedure similar to that described in Section~5 of
\citet{bakos:2007b}. The resulting bisector spans, plotted in
\reffigl{rvbis}, show no significant variation, and are not correlated
with the RVs. The combination of no detectable bisector span
variations, RV variations consistent with a Keplerian orbit in phase
with the photometric ephemeris, and short ingress and egress durations
relative to the full transit duration (\refsecl{phot}), is strong
evidence that \hatcur{} is not a blend, but rather a real TEP system.

In the same figure we also show the relative $S$ index
(\cite{vaughan:1978}), which is a measure of the chromospheric
activity of the star derived from the flux in the cores of the CaII H
and K lines. Note that our relative $S$ index has not been calibrated
to the scale of \citet{vaughan:1978}. Following \citet{isaacson:2010}, we
measured $S_{\rm HK} = 0.144$ on the calibrated scale
using the Keck/HIRES iodine-free observation. This corresponds to a
bolometric flux-corrected activity measure of $\log R^{\prime}_{\rm HK}
= -5.124$ using the definition of
\citet{noyes:1984}. The star is inactive, consistent with its slow
projected rotation velocity.

\begin{longtable}{lrrrrrr}
  \caption{Relative radial velocities, bisector spans, and activity index
	measurements of \hatcur{} from HDS/Subaru.}\label{tab:rvs}
  \hline              

  \noalign{\vskip 2pt}
  \multicolumn{1}{c}{BJD$^{\rm a}$} & 
  \multicolumn{1}{c}{RV$^{\rm b}$} & 
  \multicolumn{1}{c}{\ensuremath{\sigma_{\rm RV}}} & 
  \multicolumn{1}{c}{BS} & 
  \multicolumn{1}{c}{\ensuremath{\sigma_{\rm BS}}} & 
  \multicolumn{1}{c}{S$^{\rm c}$} & 
  \multicolumn{1}{c}{Phase} \\
  \multicolumn{1}{c}{\hbox{(2,454,000$+$)}} &
  \multicolumn{1}{c}{(\ms)} &
  \multicolumn{1}{c}{(\ms)} &
  \multicolumn{1}{c}{(\ms)} &
  \multicolumn{1}{c}{(\ms)} &
  &
  \\
  [1.0ex]
  \endfirsthead
  \hline
\endhead
  \hline
\endfoot
  \hline
\multicolumn{7}{p{10cm}}{\footnotesize $^{\rm a}$Barycentric Julian dates throughout the paper are calculated from Coordinated Universal Time (UTC).} \\
\multicolumn{7}{p{10cm}}{\footnotesize $^{\rm b}$The zero-point of these velocities is arbitrary. An overall offset $\gamma_{\rm rel}$ fitted to these velocities in \refsecl{globmod} has {\em not} been subtracted.} \\
\multicolumn{7}{p{10cm}}{\footnotesize $^{\rm c}$Relative chromospheric activity index, not calibrated to the scale of \citet{vaughan:1978}.} \\
\multicolumn{7}{p{10cm}}{\footnotesize For the iodine-free template exposures there is no RV measurement, but the BS and S index can still be determined.} \\
\endlastfoot
  \hline
    \noalign{\vskip 2pt}
    $ 1780.01380 $ & $   -21.74 $ & $     8.25 $ & $    16.23 $ & $     5.83 $ & $    0.0035 $ & $   0.091 $\\
$ 1780.02157 $ & $   -13.63 $ & $     7.70 $ & $     8.53 $ & $     6.48 $ & $    0.0037 $ & $   0.092 $\\
$ 1780.02933 $ & $   -19.59 $ & $     7.38 $ & $     0.36 $ & $     7.53 $ & $    0.0036 $ & $   0.094 $\\
$ 1780.04091 $ & $\cdots$      & $\cdots$      & $    -3.86 $ & $     7.37 $ & $    0.0033 $ & $   0.097 $\\
$ 1780.05562 $ & $\cdots$      & $\cdots$      & $    -4.85 $ & $     7.56 $ & $    0.0038 $ & $   0.100 $\\
$ 1781.02601 $ & $   -29.96 $ & $     4.50 $ & $   -10.30 $ & $     7.76 $ & $    0.0035 $ & $   0.309 $\\
$ 1781.04075 $ & $   -37.24 $ & $     4.80 $ & $    -4.91 $ & $     7.07 $ & $    0.0025 $ & $   0.312 $\\
$ 1782.01133 $ & $   -10.45 $ & $     8.27 $ & $     4.49 $ & $     7.05 $ & $    0.0038 $ & $   0.521 $\\
$ 1782.02606 $ & $    12.66 $ & $     5.81 $ & $    -4.07 $ & $     7.29 $ & $    0.0040 $ & $   0.524 $\\
$ 1782.04078 $ & $     1.68 $ & $     6.12 $ & $    -4.23 $ & $     7.21 $ & $    0.0034 $ & $   0.528 $\\
$ 1782.11167 $ & $    13.24 $ & $     6.10 $ & $     3.90 $ & $     6.88 $ & $    0.0033 $ & $   0.543 $\\
$ 1782.12638 $ & $    12.33 $ & $     6.07 $ & $     0.10 $ & $     7.83 $ & $    0.0034 $ & $   0.546 $\\
$ 1783.03706 $ & $    28.32 $ & $     6.61 $ & $    -5.54 $ & $     6.88 $ & $    0.0043 $ & $   0.742 $\\
$ 1783.05178 $ & $    35.64 $ & $     5.88 $ & $    -3.03 $ & $     6.83 $ & $    0.0034 $ & $   0.745 $\\
$ 1783.06650 $ & $    39.23 $ & $     5.27 $ & $     3.16 $ & $     6.89 $ & $    0.0035 $ & $   0.749 $\\
$ 1783.08123 $ & $    33.89 $ & $     5.84 $ & $     4.03 $ & $     6.82 $ & $    0.0034 $ & $   0.752 $\\

    [1.0ex] 
\end{longtable}

\subsection{Photometric follow-up observations}
\label{sec:phot}

\begin{figure}
  \begin{center}
    \FigureFile(80mm,112mm){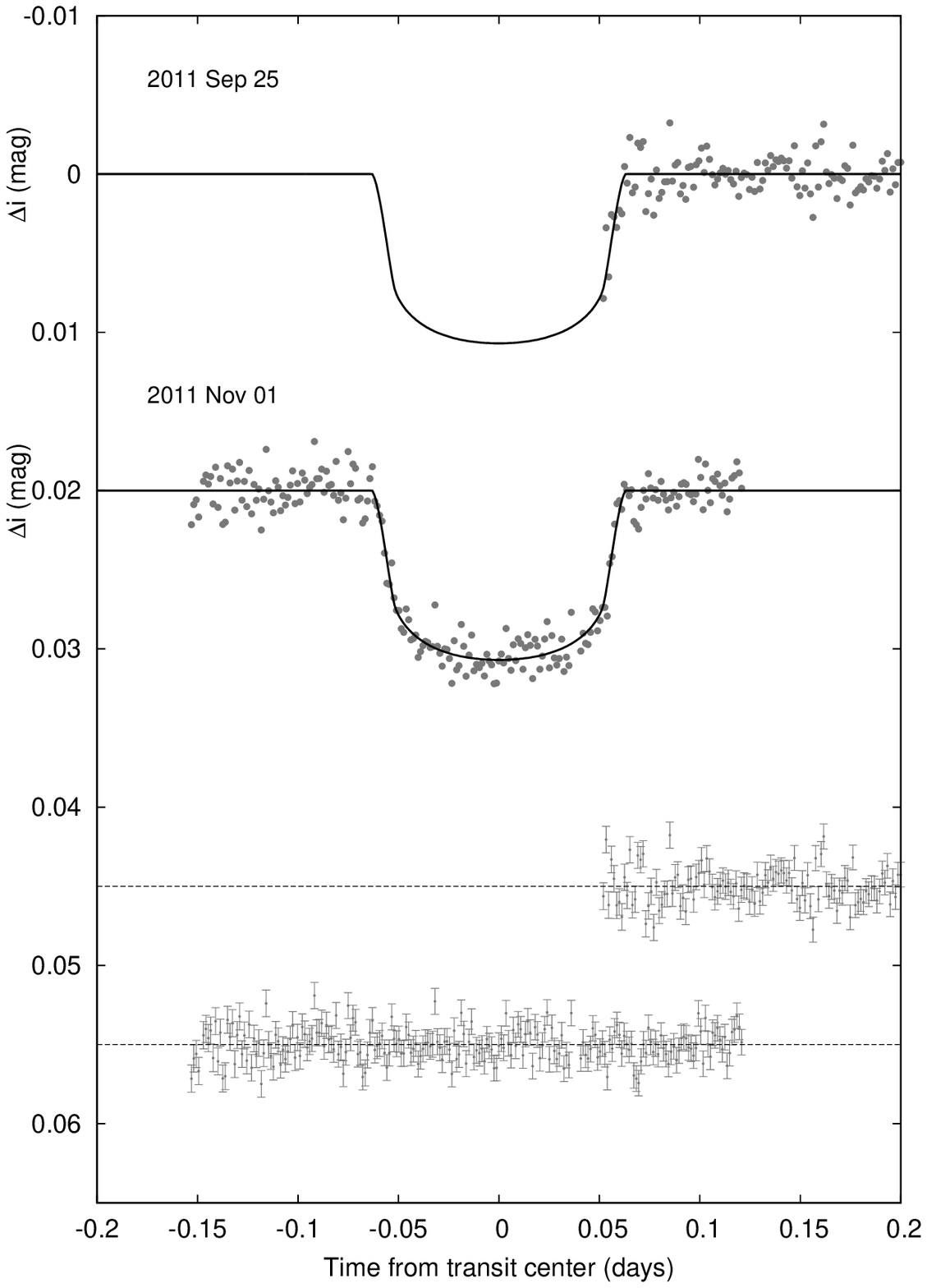}
  \end{center}
  \caption{
	Unbinned instrumental \band{i} transit \lcs{},
        acquired with KeplerCam at the \flwof{} telescope. The light
        curves have been EPD- and TFA-processed, as described in
        \refsec{globmod}.
    The dates of the events are indicated.  Curves after the first are
    displaced vertically for clarity.  Our best fit from the global
    modeling described in \refsecl{globmod} is shown by the solid
    lines.  Residuals from the fits are displayed at the bottom, in the
    same order as the top curves.  The error bars represent the photon
    and background shot noise, plus the readout noise.
}\label{fig:lc}
\end{figure}

In order to accurately characterize the system, we conducted
additional photometric observations with the KeplerCam CCD camera on
the \flwof{} telescope. Observations were performed on the nights of
2011 September 25, and 2011 November 1, in both cases using a Sloan
$i$ filter. On the first night we used an exposure time of
100\,s, on the second night we used an exposure time of 90\,s.
Due to the limited accuracy of the ephemeris calculated from the
HATNet data the transit occurred earlier than
expected on 2011 September 25, so that only the end of egress was
observed. We used this data to revise the ephemeris so that a full
transit was observed on the night of 2011 November 01.

The reduction of these images, including basic calibration,
astrometry, and aperture photometry, was performed in a similar
fashion to previous HATNet discoveries (\cite{bakos:2010}).  We used
the ``ELTG'' model described in \citet{bakos:2010} to remove trends
simultaneously with the light curve modeling.  The final time series
are shown in the top portion of \reffigl{lc}, along with our best-fit
transit \lc{} model described below; the individual measurements are
reported in \reftabl{phfu}.


\begin{longtable}{lllll}
  \caption{Differential photometry of \hatcur}\label{tab:phfu}
  \hline              
  \noalign{\vskip 2pt}
  \multicolumn{1}{c}{BJD} &
  \multicolumn{1}{c}{Mag$^{\rm a}$} &
  \multicolumn{1}{c}{\ensuremath{\sigma_{\rm Mag}}} &
  \multicolumn{1}{c}{Mag(orig)$^{\rm b}$} &
  \multicolumn{1}{c}{Filter} \\
  \multicolumn{1}{c}{\hbox{(2,400,000$+$)}} &
  &
  &
  &
  \\
  [1.0ex]
\endfirsthead
  \hline
  \\
\endhead
  \hline
\endfoot
  \hline
\multicolumn{5}{p{7cm}}{\footnotesize $^{\rm a}$The out-of-transit level has been subtracted. These magnitudes have
	been processed with the EPD and TFA procedures.} \\
\multicolumn{5}{p{7cm}}{\footnotesize $^{\rm b}$Raw magnitude values without application of the EPD and TFA
	procedures.} \\
\multicolumn{5}{p{7cm}}{\footnotesize This table is available in a machine-readable form in the online journal.  A portion is shown here for guidance regarding its form
    and content.} \\
\endlastfoot
  \hline
\noalign{\vskip 2pt}
$ 55830.68887 $ & $   0.00786 $ & $   0.00085 $ & $  11.10930 $ & $ i$\\
$ 55830.69021 $ & $   0.00339 $ & $   0.00084 $ & $  11.10510 $ & $ i$\\
$ 55830.69152 $ & $   0.00650 $ & $   0.00084 $ & $  11.10810 $ & $ i$\\
$ 55830.69282 $ & $   0.00256 $ & $   0.00084 $ & $  11.10410 $ & $ i$\\
$ 55830.69414 $ & $   0.00271 $ & $   0.00084 $ & $  11.10430 $ & $ i$\\
$ 55830.69546 $ & $   0.00338 $ & $   0.00084 $ & $  11.10500 $ & $ i$\\
$ 55830.69678 $ & $   0.00228 $ & $   0.00084 $ & $  11.10360 $ & $ i$\\
$ 55830.69808 $ & $   0.00251 $ & $   0.00084 $ & $  11.10380 $ & $ i$\\
$ 55830.69939 $ & $  -0.00047 $ & $   0.00083 $ & $  11.10140 $ & $ i$\\
$ 55830.70071 $ & $   0.00058 $ & $   0.00083 $ & $  11.10240 $ & $ i$\\

 [1.0ex]
\end{longtable}

\section{Analysis}
\label{sec:analysis}

The analysis of the \hatcur{} system, including determinations of the
properties of the host star and planet, was carried out in a similar
fashion to previous HATNet discoveries. Below we
briefly summarize the procedure and the results for \hatcur{}.

\subsection{Properties of the parent star}
\label{sec:stelparam}

Stellar atmospheric parameters were measured using our template
spectrum obtained with the Keck/HIRES instrument, and the analysis
package known as Spectroscopy Made Easy (SME; \cite{valenti:1996}),
along with the atomic line database of \citet{valenti:2005}.  SME
yielded the following values and uncertainties:
effective temperature $\teffstar=\hatcurSMEiteff$\,K, 
metallicity $\feh=\hatcurSMEizfeh$\,dex, 
stellar surface gravity $\loggstar=\hatcurSMEilogg$\,(cgs),
and projected rotation velocity $\vsini=\hatcurSMEivsin\,\kms$.

The values of \teffstar, \loggstar, and \feh\ were used to determine
the limb-darkening coefficients needed in the global modeling of the
follow-up photometry.  Following \citet{sozzetti:2007} we used \arstar, the
normalized semimajor axis, rather than the surface gravity estimated
from the spectrum, in determining the physical stellar parameters. We
combined \arstar, \teffstar, and \feh\ with stellar evolution models
from the \hatcurisofull\ series by \citet{\hatcurisocite} to determine
probability distributions of other stellar properties, including
\loggstar.
The result for the surface gravity, $\loggstar = \hatcurISOlogg$, is
consistent with the value from our SME analysis, so we did not perform
an additional iteration of SME with a fixed surface gravity. 

\begin{figure}
  \begin{center}
    \FigureFile(80mm,56mm){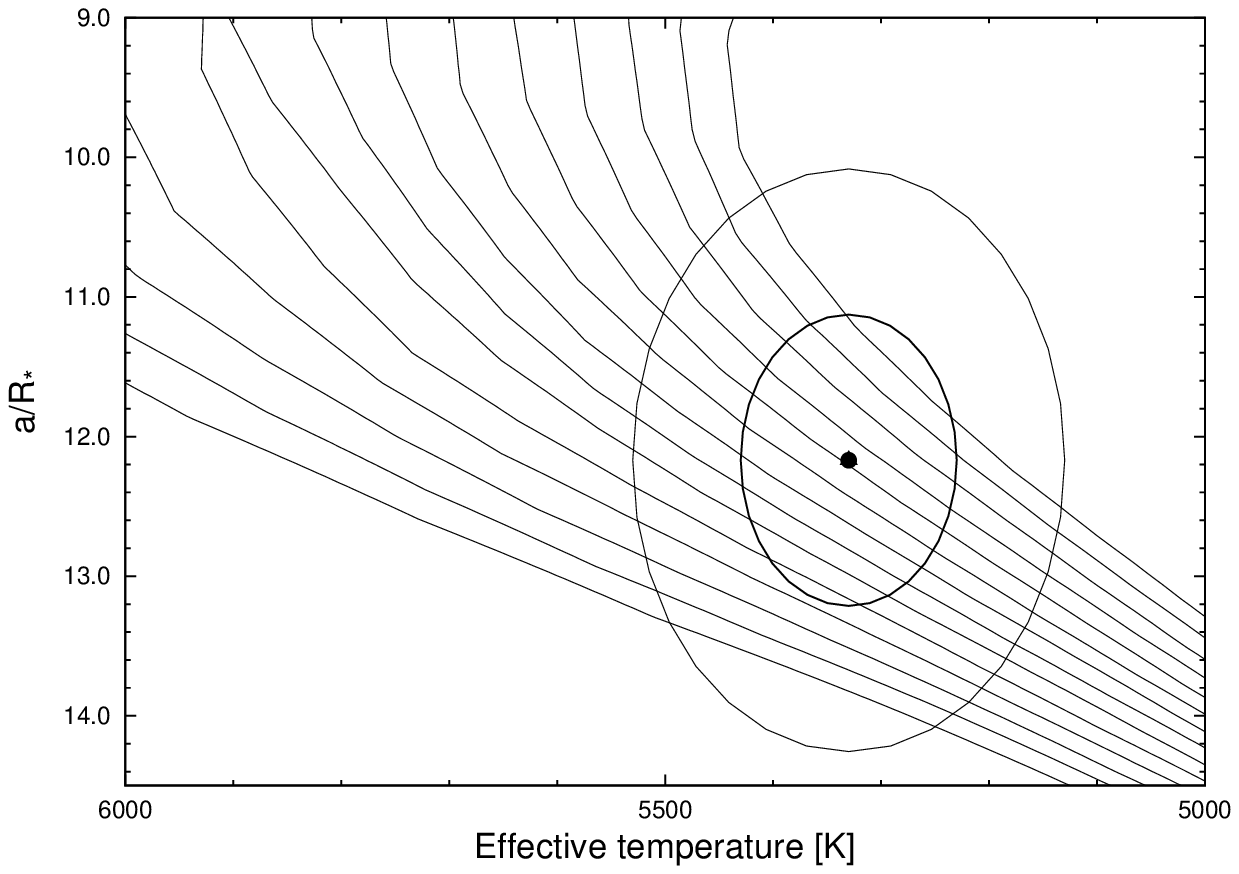}
  \end{center}
  \caption{
	Model isochrones from \citet{\hatcurisocite} for the measured
        metallicity of \hatcur, \feh = \hatcurSMEiizfehshort, and ages
        of 1 to 14\,Gyr in steps of 1\,Gyr (left to right).  The
        adopted values of $\teffstar$ and \arstar\ are shown together
        with their 1-$\sigma$ and 2-$\sigma$ confidence ellipsoids.
}\label{fig:iso}
\end{figure}

We find that \hatcur{} has a stellar mass and radius of \mstar\ =
\hatcurISOmlong\,\msun\ and \rstar\ = \hatcurISOrlong\,\rsun,
respectively. These, along with other properties, are listed in
\reftabl{stellar}.  We find that the system has an age greater than
$1.7$\,Gyr with $95\%$ confidence.  We show the inferred location of
the star in a diagram of \arstar\ versus \teffstar, analogous to the
classical H-R diagram, in Figure~\ref{fig:iso}.



\begin{longtable}{lll}
  \caption{	Stellar parameters for \hatcur{}}
	\label{tab:stellar}
  \hline              
  \noalign{\vskip 2pt}
  Parameter &
  Value &
  Source \\
\endfirsthead
  \hline
\endhead
  \hline
\endfoot
  \hline
\multicolumn{3}{p{7cm}}{\footnotesize $^{\rm a}$SME = ``Spectroscopy Made Easy'' package for the analysis of
	high-resolution spectra (\cite{valenti:1996}).} \\

\multicolumn{3}{p{7cm}}{\footnotesize $^{\rm b}$\hatcurisoshort+\hatcurlumind+SME = Based on the \hatcurisoshort\
    isochrones (\cite{\hatcurisocite}), \hatcurlumind\ as a luminosity
    indicator, and the SME results.} \\
\multicolumn{3}{p{7cm}}{\footnotesize $^{\rm c}$Determined by comparing the 2MASS $K_{s}$ magnitude to the predicted absolute $M_{K}$ magnitude.} \\
\endlastfoot
  \hline
  \noalign{\vskip 2pt}
\multicolumn{3}{l}{{\bf Spectroscopic properties}} \\
~~~~$\teffstar$ (K)\dotfill         &  \hatcurSMEteff       & SME$^{\rm a}$\\
~~~~$\feh$\dotfill                  &  \hatcurSMEzfeh       & SME                 \\
~~~~$\vsini$ (\kms)\dotfill         &  \hatcurSMEvsin       & SME                 \\
~~~~$\vmac$ (\kms)\dotfill          &  \hatcurSMEvmac       & SME                 \\
~~~~$\vmic$ (\kms)\dotfill          &  \hatcurSMEvmic       & SME                 \\
~~~~$\gamma_{\rm RV}$ (\kms)\dotfill&  \hatcurTRESgamma       & TRES              \\
~~~~$\log R^{\prime}_{\rm HK}$\dotfill& $-5.124$             & HIRES               \\
\multicolumn{3}{l}{{\bf Photometric properties}} \\
~~~~$V$ (mag)\dotfill               &  \hatcurCCtassmv      & TASS                \\
~~~~$V\!-\!I_C$ (mag)\dotfill       &  \hatcurCCtassvi      & TASS                \\
~~~~$J$ (mag)\dotfill               &  \hatcurCCtwomassJmag & 2MASS           \\
~~~~$H$ (mag)\dotfill               &  \hatcurCCtwomassHmag & 2MASS           \\
~~~~$K_s$ (mag)\dotfill             &  \hatcurCCtwomassKmag & 2MASS           \\
\multicolumn{3}{l}{{\bf Derived properties}} \\
~~~~$\mstar$ ($\msun$)\dotfill      &  \hatcurISOmlong      & \hatcurisoshort+\hatcurlumind+SME$^{\rm b}$\\
~~~~$\rstar$ ($\rsun$)\dotfill      &  \hatcurISOrlong      & \hatcurisoshort+\hatcurlumind+SME         \\
~~~~$\loggstar$ (cgs)\dotfill       &  \hatcurISOlogg       & \hatcurisoshort+\hatcurlumind+SME         \\
~~~~$\lstar$ ($\lsun$)\dotfill      &  \hatcurISOlum        & \hatcurisoshort+\hatcurlumind+SME         \\
~~~~$M_V$ (mag)\dotfill             &  \hatcurISOmv         & \hatcurisoshort+\hatcurlumind+SME         \\
~~~~$M_K$ (mag,\hatcurjhkfilset)\dotfill &  \hatcurISOMK    & \hatcurisoshort+\hatcurlumind+SME         \\
~~~~Age (Gyr)\dotfill               &  \hatcurISOage        & \hatcurisoshort+\hatcurlumind+SME         \\
~~~~Distance (pc)$^{\rm c}$\dotfill           &  \hatcurXdist         & \hatcurisoshort+\hatcurlumind+SME\\
\end{longtable}

\subsection{Global modeling of the data}
\label{sec:globmod}

We modeled the HATNet photometry, the follow-up photometry, and the
high-precision RV measurements using the procedure described in detail
by \citet{bakos:2010}. 

The resulting parameters pertaining to the light curves and RV curves,
together with derived physical parameters of the planet, are listed in
\reftabl{planetparam}.

We find a mass for the planet of
$\mpl=\hatcurPPmlong\,\mjup$ and a radius of
$\rpl=\hatcurPPrlong\,\rjup$, leading to a mean density
$\rho_p=\hatcurPPrho$\,\gcmc. 
These and other planetary parameters are listed at the bottom of
Table~\ref{tab:planetparam}.
We note that the eccentricity of the orbit is consistent with zero: 
$k \equiv e\cos(\omega) = \hatcurRVk$ and 
$h \equiv e\sin(\omega) = \hatcurRVh$ 
($e = \hatcurRVeccen$, $\omega = \hatcurRVomega\arcdeg$). We find $e <
0.165$ with 95\% confidence.

\begin{longtable}{ll}
  \caption{Orbital and planetary parameters}\label{tab:planetparam}
  \noalign{\vskip 2pt}
  \hline              
	~~~~~~~~~~~~~~~~~~Parameter~~~~~~~~~~~~~~~~~~ &
	Value \\
  [1.0ex]
\endfirsthead
  \hline
\endhead
  \hline
\endfoot
  \hline
\multicolumn{2}{p{7cm}}{\footnotesize $^{\rm a}$    \ensuremath{T_c}: Reference epoch of mid transit that minimizes the
    correlation with the orbital period. BJD is calculated from UTC.
	\ensuremath{T_{14}}: total transit duration, time between first to
	last contact;
	\ensuremath{T_{12}=T_{34}}: ingress/egress time, time between first
	and second, or third and fourth contact.
} \\
\multicolumn{2}{p{7cm}}{\footnotesize $^{\rm b}$ The reciprocal of the half duration of the transit, related to \arstar, is used as a jump parameter in the MCMC procedure.} \\
\multicolumn{2}{p{7cm}}{\footnotesize $^{\rm c}$
	Values for a quadratic law, adopted from the tabulations by
    \citet{claret:2004} according to the spectroscopic (SME) parameters
    listed in \reftabl{stellar}.
} \\
\multicolumn{2}{p{7cm}}{\footnotesize $^{\rm d}$
    $k_{\rm RV} = e\cos\omega$ and $h_{\rm RV} = e\sin\omega$. These
  are determined from the global modeling, and primarily constrained
  by the RV data.
} \\
\multicolumn{2}{p{7cm}}{\footnotesize $^{\rm e}$
	Correlation coefficient between the planetary mass \mpl\ and radius
	\rpl.
} \\
\multicolumn{2}{p{7cm}}{\footnotesize $^{\rm f}$
	The Safronov number is given by $\Theta = \frac{1}{2}(V_{\rm
	esc}/V_{\rm orb})^2 = (a/\rpl)(\mpl / \mstar )$
	(see \cite{hansen:2007}).
} \\
\multicolumn{2}{p{7cm}}{\footnotesize $^{\rm g}$
	Incoming flux per unit surface area, averaged over the orbit.
} \\
\endlastfoot
  \hline
  \noalign{\vskip 2pt}
\multicolumn{2}{l}{{\bf \Lc{} parameters}} \\
~~~$P$ (days)             \dotfill    & $\hatcurLCP$              \\
~~~$T_c$ (${\rm BJD}$)$^{\rm a}$    
      \dotfill    & $\hatcurLCT$              \\
~~~$T_{14}$ (days)$^{\rm a}$
         \dotfill    & $\hatcurLCdur$            \\
~~~$T_{12} = T_{34}$ (days)$^{\rm a}$
         \dotfill    & $\hatcurLCingdur$         \\
~~~$\arstar$              \dotfill    & $\hatcurPPar$             \\
~~~$\zrstar$$^{\rm b}$              \dotfill    & $\hatcurLCzeta$           \\
~~~$\rpl/\rstar$          \dotfill    & $\hatcurLCrprstar$        \\
~~~$b^2$                  \dotfill    & $\hatcurLCbsq$            \\
~~~$b \equiv a \cos i/\rstar$
                          \dotfill    & $\hatcurLCimp$            \\
~~~$i$ (deg)              \dotfill    & $\hatcurPPi$              \\

\multicolumn{2}{l}{{\bf Limb-darkening coefficients}$^{\rm c}$} \\
~~~$a_i$ (linear term, $i$ filter)    \dotfill    & $\hatcurLBii$             \\
~~~$b_i$ (quadratic term) \dotfill    & $\hatcurLBiii$            \\

\multicolumn{2}{l}{{\bf RV parameters}} \\
~~~$K$ (\ms)              \dotfill    & $\hatcurRVK$              \\
~~~$k_{\rm RV}$$^{\rm d}$ 
                          \dotfill    & $\hatcurRVk$              \\
~~~$h_{\rm RV}$$^{\rm d}$
                          \dotfill    & $\hatcurRVh$              \\
~~~$e$                    \dotfill    & $\hatcurRVeccen$          \\
~~~$\omega$ (deg)         \dotfill    & $\hatcurRVomega$          \\
~~~RV fit RMS (\ms)        \dotfill    & \hatcurRVfitrms           \\

\multicolumn{2}{l}{{\bf Secondary eclipse parameters}} \\
~~~$T_s$ (BJD)            \dotfill    & $\hatcurXsecondary$       \\
~~~$T_{s,14}$             \dotfill    & $\hatcurXsecdur$          \\
~~~$T_{s,12}$             \dotfill    & $\hatcurXsecingdur$       \\

\multicolumn{2}{l}{{\bf Planetary parameters}} \\
~~~$\mpl$ ($\mjup$)       \dotfill    & $\hatcurPPmlong$          \\
~~~$\rpl$ ($\rjup$)       \dotfill    & $\hatcurPPrlong$          \\
~~~$C(\mpl,\rpl)$$^{\rm e}$
         \dotfill    & $\hatcurPPmrcorr$         \\
~~~$\rhopl$ (\gcmc)       \dotfill    & $\hatcurPPrho$            \\
~~~$\log g_p$ (cgs)       \dotfill    & $\hatcurPPlogg$           \\
~~~$a$ (AU)               \dotfill    & $\hatcurPParel$           \\
~~~$T_{\rm eq}$ (K)       \dotfill    & $\hatcurPPteff$           \\
~~~$\Theta$$^{\rm f}$\dotfill  & $\hatcurPPtheta$          \\
~~~$\langle F \rangle$ ($10^{\hatcurPPfluxavgdim}$\ergscmsq)$^{\rm g}$ 
         \dotfill    & $\hatcurPPfluxavg$        \\
[1.0ex]
\end{longtable}


\section{Discussion}
\label{sec:discussion}

\begin{figure}
  \begin{center}
    \FigureFile(80mm,80mm){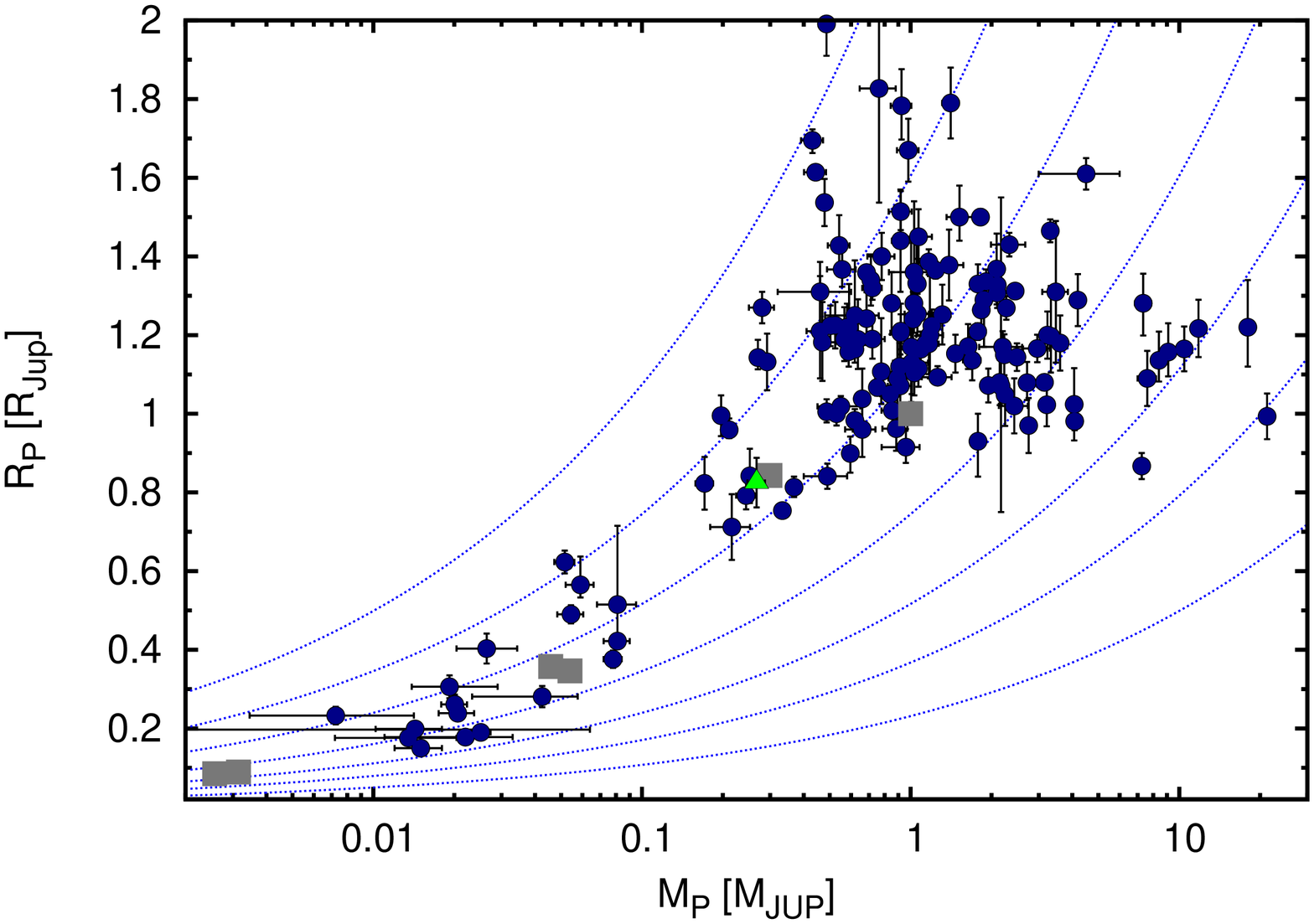}
  \end{center}
  \caption{
    Mass--radius diagram of known TEPs (blue circles).
    \hatcurb\ is shown as a green triangle.  Overlaid are
    isodensity lines for 0.1, 0.3, 0.9, 3.0, 9.0, 25.0, and 100\,\gcmc
    (dotted lines). Solar system planets are shown with gray
    squares.
}\label{fig:exomr}
\end{figure}

\hatcurb{} belongs to the small, but increasing group of Saturn-mass
planets ($0.1 < M < 0.4 M_J$; Figure~\ref{fig:exomr}; Table~\ref{tab:saturn}).
Of the 12 known Saturn-mass TEPs,
\hatcurb{} is the planet closest in mass and radius to Saturn, along
with WASP-29b ($\mpl=0.244\pm0.020~\mjup$,
$\rpl=0.792^{+0.056}_{-0.035}~\rjup$, $\rhopl=0.65\pm0.10~\gcmc$;
\cite{hellier:2010}) and the somewhat lower density planets Kepler-9b
and Kepler-9c ($\rhopl\sim$0.4-0.5 $\gcmc$; \cite{holman:2010}).
Other Saturn-mass TEPs include the inflated, low-density
($\rhopl\sim$0.3 $\gcmc$) planets HAT-P-12b (\cite{hartman:2009}),
HAT-P-18b, HAT-P-19b (\cite{hartman:2011a}), WASP-21b
(\cite{bouchy:2010}), and WASP-39b (\cite{faedi:2011}), and the dense
planets CoRoT-8b (\cite{borde:2010}), Kepler-16ABb
(\cite{doyle:2011}), and HD 149026b (\cite{sato:2005}) with
$\rhopl\gtrsim$0.9 $\gcmc$.

For 0.1--0.6 $\mjup$ planets, \citet{enoch:2011} find that planetary
radius is proportional to planetary equilibrium temperature and
inversely proportional to host star metallicity, with hotter planets
having more inflated radii and planets orbiting lower metallicity
stars have larger radii (perhaps because they have smaller metal-rich
cores). The observed radius for \hatcurb{} is slightly smaller than
that expected from the relationship ($0.83$\,$\rjup$ compared to the
expected value of 0.97\,$\rjup$ based on the equation (3) in
\cite{enoch:2011}), but is still consistent with the general trend.

\hatcurb{} provides a clear example that irradiation is not the sole
factor affecting the radii of Saturn-mass planets as it has a higher
planetary equilibrium temperature ($T_{\rm eq}=1082$ K) than that of
inflated planets such as HAT-P-12b ($\rpl=0.959~\rjup$, $T_{\rm
  eq}=963$ K) and HAT-P-18b ($\rpl=0.995~\rjup$, $T_{\rm eq}=852$
K). Metallicity is clearly another important factor in setting the
radii of planets in this mass range. From the theoretical models of
\citet{fortney:2007}, \hatcurb{} could have a heavy-element core of
$\sim 25~M_{\oplus}$, fitting the general trend that
higher-metallicity systems have larger inferred cores and thus smaller
planetary radii. However, the recently discovered planets violate this
pattern; HAT-P-18b and HAT-P-19b have low densities even with high
metallicity ($\feh=+0.10$ and $+0.23$, respectively), while the
circumbinary planet Kepler-16ABb has a high density even with low
metallicity ($\rm{[m/H]}=-0.3$), suggesting that the relationship is
not so simple.  Continued discoveries over a wide range of planetary
and stellar parameters will allow us to further investigate the nature
of these Saturn-mass planets.

\begin{longtable}{ccccccc}
  \caption{Saturn-mass TEPs ($0.1 < M < 0.4 M_J$)}\label{tab:saturn}
  \hline              

  \noalign{\vskip 2pt}
  \multicolumn{1}{c}{Planet Name} & 
  \multicolumn{1}{c}{Mass} & 
  \multicolumn{1}{c}{Radius} &
  \multicolumn{1}{c}{Mean Density} & 
  \multicolumn{1}{c}{$T_{\rm eq}$$^{\rm a}$} &
  \multicolumn{1}{c}{$\feh$} &
  \multicolumn{1}{c}{Reference}\\
  \multicolumn{1}{c}{}  &
  \multicolumn{1}{c}{($\mjup$)} &
  \multicolumn{1}{c}{($\rjup$)} &
  \multicolumn{1}{c}{(\gcmc)} &
  \multicolumn{1}{c}{(K)} &
  \multicolumn{1}{c}{} &
  \multicolumn{1}{c}{}\\
  [1.0ex]
  \endfirsthead
  \hline
\endhead
  \hline
\endfoot
  \hline
\multicolumn{3}{p{7cm}}{\footnotesize $^{\rm a}$Planetary Bond albedo $A=0$ is
assumed.} \\
\multicolumn{3}{p{7cm}}{\footnotesize $^{\rm b}$The value is not $\feh$ but [m/H].} \\
\endlastfoot
  \hline
    \noalign{\vskip 2pt}
Kepler-9c   & 0.171$\pm$0.013 & 0.823$\pm$0.067  & 0.383$\pm$0.098 & 620$\pm$30 & -- & \citet{holman:2010}\\
            &      &       &      &     &    & \citet{havel:2011}\\
HAT-P-18b   & 0.197$\pm$0.013 & 0.995$\pm$0.052  & 0.25$\pm$0.04 & 852$\pm$28 & $+$0.10$\pm$0.08 & \citet{hartman:2011a}\\
HAT-P-12b   & 0.211$\pm$0.012 & 0.959$^{+0.029}_{-0.021}$  & 0.295$\pm$0.025 & 963$\pm$16 & $-$0.29$\pm$0.05 & \citet{hartman:2009}\\
CoRoT-8b    & 0.22$\pm$0.03 & 0.57$\pm$0.02  & 1.6$\pm$0.1 & 860$\pm$20 & $+$0.3$\pm$0.1$^{\rm b}$ &\citet{borde:2010}\\
WASP-29b    & 0.244$\pm$0.020 & 0.792$^{+0.056}_{-0.035}$  & 0.65$\pm$0.10 & 980$\pm$40 & $+$0.11$\pm$0.14 & \citet{hellier:2010}\\
Kepler-9b   & 0.252$\pm$0.013 & 0.842$\pm$0.069  & 0.524$\pm$0.132  & 780$\pm$30 & -- & \citet{holman:2010}\\
            &      &       &      &     &    & \citet{havel:2011}\\
HAT-P-38b   & 0.267$\pm$0.020 & 0.825$^{+0.092}_{-0.063}$  & 0.59$\pm$0.16  & 1082$\pm$55 & $+$0.06$\pm$0.1 & This work\\
WASP-39b    & 0.28$\pm$0.03 & 1.27$\pm$0.04  & 0.19$\pm$0.03 & 1116$^{+33}_{-32}$ & $-$0.12$\pm$0.10 & \citet{faedi:2011}\\
HAT-P-19b   & 0.292$\pm$0.018 & 1.132$\pm$0.072  & 0.25$\pm$0.04 & 1010$\pm$42 & $+$0.23$\pm$0.08 & \citet{hartman:2011a}\\
WASP-21b    & 0.300$\pm$0.011 & 1.07$\pm$0.06  & 0.32$\pm$0.07 & 1260$\pm$30 & $-$0.46$\pm$0.11 &\citet{bouchy:2010}\\
Kepler-16ABb& 0.333$\pm$0.016 & 0.7538$^{+0.0026}_{-0.0023}$ & 0.964$^{+0.047}_{-0.046}$ & 206$\pm$7 & $-$0.3$\pm$0.2$^{\rm b}$ &\citet{doyle:2011}\\
HD149026b & 0.368$^{+0.013}_{-0.014}$ & 0.813$^{+0.027}_{-0.025}$ & 0.85$^{+0.10}_{-0.09}$ & 1740$\pm$60 & $+$0.36$\pm$0.05 & \citet{sato:2005}\\
            &      &       &      &     &    & \citet{carter:2009}\\
    [1.0ex] 
\end{longtable}

\bigskip

HATNet operations have been funded by NASA grants NNG04GN74G,
NNX08AF23G, NNX09AB29G and SAO IR\&D grants.  
GT acknowledges partial support from NASA grant NNX09AF59G.  We
acknowledge partial support also from the Kepler Mission under NASA
Cooperative Agreement NCC2-1390 (D.W.L., PI).  G.K.~thanks the
Hungarian Scientific Research Foundation (OTKA) for support through
grant K-81373.  This research has made use of Keck telescope time
granted through NASA (N108Hr). Based in part on data collected at
Subaru Telescope, which is operated by the National Astronomical
Observatory of Japan. 






\end{document}